\documentclass[twocolumn,showpacs,showkeys,amsmath,amssymb]{revtex4}

\usepackage{graphicx}% Include figure files
\usepackage{dcolumn}% Align table columns on decimal point
\usepackage{bm}% bold math

\begin{document}

\title{Self-Similar Random Processes and Infinite-Dimensional \\
Configuration Spaces}

\author{Gerald A. Goldin}
\email{gagoldin@dimacs.rutgers.edu}
\affiliation{%
Departments of Mathematics and Physics,
Rutgers University,\\118 Frelinghuysen Road, SERC 239, Busch Campus,\\
Piscataway, NJ 08854 USA
}%
\author{Ugo Moschella}%
\email{ugo.moschella@uninsubria.it}

\affiliation{Dipartimento di Scienze Chimiche, Fisiche e Matematiche,\\
Universit\`a dell'Insubria, \\ 22100 Como, Italy \\
and INFN, Sez. di Milano, Italy}
\author{Takao Sakuraba}
\email{tsakuraba@optonline.net}
\affiliation{Department of Mathematics, Rutgers University,\\
Piscataway, NJ 08854 USA}

\begin{abstract}
We discuss various infinite-dimensional configuration spaces that
carry measures quasiinvariant under compactly-supported
diffeomorphisms of a manifold $M$ corresponding to a physical
space. Such measures allow the construction of unitary
representations of the diffeomorphism group, which are important
to nonrelativistic quantum statistical physics and to the quantum
theory of extended objects in $M = \mathbb{R}^d$. Special
attention is given to measurable structure and topology underlying
measures on generalized configuration spaces obtained from
self-similar random processes (both for $d = 1$ and $d > 1)$,
which describe infinite point configurations having accumulation
points.
\end{abstract}


%Use showkeys class option if keyword
\maketitle

\section{\label{intro}Introduction}

Let $M$ be the manifold of physical space, usually taken to be
$d$-dimensional Euclidean space $\mathbb{R}^d$. Let {\it
Diff\/}$^{\,c}(M)$ be the (infinite-dimensional) group of
compactly-supported diffeomorphisms of $M$, under composition. The
local current algebra approach to nonrelativistic quantum
mechanics led to the understanding that a wide variety of quantum
systems could be described by constructing the continuous unitary
representations (CURs) of {\it Diff\/}$^{\,c}(M)$, the group of
compactly supported diffeomorphisms of $M$ (under composition)
\cite{GolSh1970, Gol1971, GGPS1974, Men1974, VGG1977, AKR1998,
AKR1999}.

To say that the diffeomorphism $\phi$ of $M$ has {\it compact
support\/} means that for all points ${\bf{x}} \in M$ that are
outside some compact (and therefore bounded) region of $M$, the
diffeomorphism acts as the identity operator: $\phi(\bf{x}) \equiv
{\bf x}$. Our convention here will be to define the group product
$\phi_1 \phi_2 = \phi_2 \circ \phi_1$, where $\circ$ denotes the
composition of $\phi_1, \phi_2 \in$ {\it Diff\/}$^{\,c}(M)$; so
that $[\phi_1\phi_2](\mathbf{x}) = \phi_2(\phi_1(\mathbf{x}))$ for
$\mathbf{x} \in M$. Thus we have a ``right action'' of the
diffeomorphism group on the manifold.

In a very general framework, the Hilbert space where the unitary
representation of {\it Diff\/}$^{\,c}(M)$ can be realized is the
space of square-integrable functions, ${\mathcal H} \,=\,
L^{2}_{\mu}(\Delta,{\mathcal W})$; where $\Delta$ is some {\it
configuration space\/} on which the diffeomorphism group naturally
acts (with a right action), $\mu$ is a measure on $\Delta$
satisfying appropriate technical conditions, ${\mathcal W}$ is an
inner product space, and the elements of $\mathcal{H}$ are
$\mu$-measurable functions $\Psi(\gamma)$ on $\Delta$ taking
values in ${\mathcal W}$. The inner product of two such functions
in $\mathcal{H}$ is given by
\begin{equation}
(\Psi_1, \Psi_2)\, =\, \int_\Delta\, \langle
{\Psi_1(\gamma)},\Psi_2(\gamma)\rangle_{\mathcal{W}}\,d\mu(\gamma)\,<
\infty, \label{innerproduct}
\end{equation}
where $\langle
{\Psi_1(\gamma)},\Psi_2(\gamma)\rangle_{\mathcal{W}}$ denotes the
inner product in $\mathcal{W}$. Then the operators $V(\phi)$
defining a CUR are given by
\begin{equation}
[V(\phi)\Psi](\gamma) \, = \, \chi_{\phi}(\gamma)\Psi(\phi \gamma)
\sqrt {\frac{{d\mu_{\phi}} }{ {d\mu}}(\gamma)}\,; \label{rep1}
\end{equation}
where $\phi \gamma$ refers to the action of the diffeomorphism
$\phi$ on $\gamma \in \Delta$, and where $\chi_{\phi}: {\mathcal
W} \rightarrow {\mathcal W}$ is a family of unitary operators
acting in ${\mathcal W}$ satisfying a certain cocycle equation
(see below).

In this article we shall consider various candidates for a
``large'' configuration space, within which different choices of
the space $\Delta$ may be situated, that permit the construction
of measures having the necessary property of quasiinvariance under
diffeomorphisms. We then focus on the generalized configuration
space $\Sigma_M$ whose elements are {\it finite or countably
infinite subsets\/} of $M$, and discuss ways of endowing it with a
$\sigma$-algebra and a topology. The results underlie the
construction of measures on generalized configuration spaces
obtained from self-similar random processes in $\mathbb{R}^d$
(both for $d = 1$ and $d > 1)$, which describe infinite point
configurations having accumulation points.

In Sec. II we briefly discuss the meaning of Eq. \!(\ref{rep1}),
reviewing the necessary concepts. Sec. III surveys some aspects of
several possible choices of ``large'' configuration spaces, while
Sec. IV focuses on topology and measurable structure in
$\Sigma_M$. In Sec. V, we give a rapid overview of the
construction of certain families of quasiinvariant measures in
$\Sigma_{\mathbb{R}^d}$ making use of self-similar random
processes.

\section{\label{measureandcocycle}Measures and Cocycles}

The measure $\mu$ that appears in Eq. \!(\ref{rep1}) and in the
definition of $\mathcal{H}$ is, as usual, a countably-additive,
positive real-valued function defined on a $\sigma$-algebra
$\mathcal{M}$ of subsets of $\Delta$. It is required to have the
key property of {\it quasiinvariance\/} under the action of
diffeomorphisms on $\Delta$.

In general, let $G$ be a group of transformations of a measurable
space $(X,{\mathcal M})$, where ${\mathcal M}$ is a $G$-invariant
${\sigma}$-algebra of subsets of $X$. A measure $\mu$ on
${\mathcal M}$ is said to be {\it invariant\/} under $G$ if and
only if for all $E \in \mathcal{M}$, and for all $g \in G$,
$\mu(gE) = \mu(E)$. It is said to be {\it quasiinvariant\/} under
$G$ if and only if for all $E \in \mathcal{M}$ such that $\mu (E)
> 0$, and for all $g \in G$, $\mu(g(E)) > 0$. That is, $g \in G$
acts on $X$ in such a way as to preserve the class of sets that
have $\mu$-measure zero.

Quasiinvariance is {\it a fortiori\/} a consequence of invariance,
but not conversely. For example, Lebesgue measure $d{\bf x}$ on $X
= \mathbb{R}^d$ is invariant under the group of rigid motions
(translations and rotations). It is quasiinvariant, but not
invariant, under the group of compactly-supported diffeomorphisms
of ${\mathbb R}^d$.

For $\phi \in G =$ {\it Diff\/}$^{\,c}(M)$ acting on $X = \Delta$,
define the transformed measure $\mu_{\phi}$ by setting
$\mu_{\phi}(E) = \mu(\phi(E))$ for any $E \in {\mathcal M}$.
Because of the group structure and the $G$-invariance of
${\mathcal M}$, the quasiinvariance of $\mu$ under $G$ is
equivalent to the absolute continuity of $\mu_{\phi_1}$ with
respect to $\mu_{\phi_2}$ for any ${\phi_1}, {\phi_2} \in G$. In
particular, the quasiinvariance of $\mu$ is necessary and
sufficient for the existence of the Radon-Nikodym (RN) derivative
$({{d\mu_{\phi}} /{d\mu}})(\gamma)$ appearing in Eq.
\!(\ref{rep1}), for all elements $\phi \in$ {\it Diff\/}$^c(M)$.
For example, with $M = \mathbb{R}^d$, $\Delta = \mathbb{R}^d$, and
$d\mu = d{\bf x}$, we have $({{d\mu_{\phi}} /{d\mu}})({\bf x}) =
\mathcal{J}_\phi(\bf{x})$, the Jacobian of $\phi$ at $\bf{x}$.
Since $\phi$ has compact support, we have
$\mathcal{J}_\phi(\bf{x}) \equiv$ $1$ outside some bounded region
of $\mathbb{R}^d$.

The square root of the RN derivative in Eq. \!(\ref{rep1}) is
precisely the factor necessary to make the operators $V(\phi)$
unitary in $\mathcal{H}$, since $\chi_\phi(\gamma)$ is to be taken
as acting unitarily in $\mathcal{W}$ (see below). That is, the
diffeomorphism $\phi$ {\it moves\/} the argument of the wave
function $\Psi$, and the square root factor {\it corrects\/} so
that when we calculate the inner product
$(V(\phi)\Psi_1,V(\phi)\Psi_2)$ using Eq. \!(\ref{innerproduct}),
we find that we have merely made the change of variable
$\gamma^{\prime} = \phi\gamma$ under the integral sign.

Let ${\mathcal D}(M)$ be the space of real-valued,
compactly-supported $C^\infty$ functions $f$ on $M$. We have then
the natural semidirect product group ${\mathcal D}(M) \times$ {\it
Diff\/}$^{\,c}(M)$, with the group law given by
\begin{equation}
(f_1, \phi_1)(f_2, \phi_2) = (f_1 +f_2\circ \phi_1, \phi_1
\phi_2). \label{grouplaw}
\end{equation}
Now it may sometimes be the case that $V(\phi)$ is a
subrepresentation of a CUR of ${\mathcal D}(M) \times$ {\it
Diff\/}$^{\,c}(M)$, which we write $U(f)V(\phi)$. Then the
operators $U(f), f \in {\mathcal D}(M)$, typically act in
${\mathcal H}$ as multiplication operators, consistently with Eq.
\!(\ref{rep1}):
\begin{equation}
[U(f)\Psi](\gamma) = \exp [i\langle \gamma, f \rangle
]\Psi(\gamma)\,,\label{repU}
\end{equation}
where $\langle \gamma, f \rangle$ denotes an action of $\gamma \in
\Delta$ on $f \in {\mathcal D}(M)$ as a continuous linear
functional. That is, the configuration $\gamma$ is here identified
with a {\it distribution,\/} and $\Delta$ is identified with a
subset of the dual space ${\mathcal D}^{\,\prime}(M)$. This is one
of the possibilities discussed in Sec. III.

In Eq. \!(\ref{rep1}), $\chi_{\phi}: {\mathcal W} \rightarrow
{\mathcal W}$ is a family of unitary operators in ${\mathcal W}$
satisfying the {\it cocycle equation\/}
\begin{equation}
\chi_{\phi_{1}}(\gamma)\chi_{\phi_{2}}(\phi_{1} \gamma) \, = \,
\chi_{\phi_{1}\phi_{2}}(\gamma)\,, \label{cocycle}
\end{equation}
which holds almost everywhere (a.e.) in $\Delta$ for each pair of
diffeomorphisms $\phi_{1}, \phi_{2}$. That is, Eq.
\!(\ref{cocycle}) holds outside a $\mu$-measure zero set that in
general may depend on $\phi_1$ and $\phi_2$.

The cocycle equation follows directly from the condition that $V$
respect the group law, $V(\phi_1)V(\phi_2) = V(\phi_1\phi_2)$. The
trivial cocycle $\chi_{\phi}(\gamma) \equiv I$ is always
permitted, and in the case of a CUR describing $N$ identical
particles, this choice corresponds to Bose-Einstein statistics.
Inequivalent choices of $\chi_{\phi}$ (noncohomologous cocycles)
are associated with Fermi-Dirac statistics, nontrivial phase
effects, and anyon statistics in two space dimensions
\cite{LeiMyr1977,GolMenSh1981,Wil1982,GolSh1983,GolMenSh1983}, as
well as with certain nonlinear variations of quantum mechanics
\cite{DoebGol1992,Gol1992,DoebGol1996}. In the simplest cases,
$\mathcal{W}$ is just the $1$-dimensional space of complex numbers
$\mathbb{C}$, so that we have complex-valued wave functions on
$\Delta$. Then the $\chi_\phi$ act through multiplication by
complex numbers of modulus $1$. Higher-dimensional choices for
$\mathcal{W}$ are associated with paraparticles in $\mathbb{R}^3$
and plektons in $\mathbb{R}^2$
\cite{Greenberg,GolMenSh1985,Gol1987}.

\section{\label{genconfspaces}General Configuration Spaces}

To this point no universal configuration space for the
representation theory of {\it Diff\/}$^{\,c}(M)$ has been agreed
upon. Consequently we have no one universal configuration space
for the physics of systems with infinitely-many degrees of freedom
in $\mathbb{R}^d$, within which specific choices of configuration
spaces for particular systems are situated. This very likely
reflects a gap in our present level of understanding. Let us
describe here some choices that have been made, that allow the
convenient description and interpretation of certain classes of
unitary representations.

\subsection{Locally finite point configurations}

The standard configuration space for statistical physics is the
space $\Gamma^{(\infty)}_M$ of countably infinite but locally
finite subsets of $M$, where usually $M = \mathbb{R}^d$.
Frequently one considers the disjoint union of this space with the
spaces of $N$-point subsets; thus $\Gamma_{M} =
\bigsqcup_{N=1}^{\infty} \Gamma_{M}^{(N)} \bigsqcup
\Gamma^{(\infty)}_{M}$ is the space of all locally finite subsets
of $M$. Measures on the configuration space
$\Gamma^{(\infty)}_{\mathbb{R}^d}$ describe equilibrium states in
$\mathbb{R}^d$ in statistical mechanics; while
$\Gamma^{(\infty)}_{\mathbb{R}^d}$ also enters quantum theory in
the description of infinite gases of quantum particles in
${\mathbb{R}^d}$.

Let $|\gamma|$ denote the cardinality of the set $\gamma$. A
configuration $\gamma \subset \mathbb{R}^d$ in
$\Gamma^{(\infty)}_{\mathbb{R}^d}$ has the properties that
$|\gamma | = \aleph_{\,0}$, while for any compact set $K \subset
\mathbb{R}^d,\,\, |\gamma \cap K | < \infty$. Then the
diffeomorphism $\,\phi \in$ {\it Diff\/}$^c(\mathbb{R}^d)\,$ acts
naturally on any configuration $\,\gamma \in
\Gamma_{\mathbb{R}^d}\,$ by its action on the individual elements
of $\gamma$. This clearly respects the property of being finite or
locally finite. Measures on $\Gamma^{(\infty)}_{\mathbb{R}^d}$
that are quasiinvariant under diffeomorphisms have been
extensively studied, and include Poisson measures and Gibbsian
measures \cite{GGPS1974, VGG1977, AKR1998, AKR1999, KK2002}.

In particular, the choice of a Poisson measure $d\mu^\sigma$ on
$\Gamma^{(\infty)}_{\mathbb{R}^d}$, with intensity $\sigma > 0$,
together with the trivial cocycle $\chi_\phi \equiv 1$, gives a
CUR of {\it Diff\/}$^c(\mathbb{R}^{d})$ via Eq. \!(\ref{rep1}).
This representation describe the infinite, free quantum Bose gas
having $\sigma$ as its average particle number density
\cite{GGPS1974}. Here we have, for any choice of $\sigma$,
\begin{equation}
\frac{d\mu^{\sigma}_{\phi}}{d\mu^\sigma}(\gamma) = \prod_{{\mathbf
x} \in \gamma} \mathcal{J}_\phi({\mathbf x})\,. \label{PoissonRN}
\end{equation}
Since $\phi$ has compact support and $\gamma$ is locally finite,
it is evident that all but a finite number of terms in the
infinite product of Jacobians in Eq. \!(\ref{PoissonRN}) are equal
to $1$. Thus this product gives a {\it finite, nonzero result\/}
for the value of the RN derivative---expressing the fact that
Poisson measures on $\Gamma^{(\infty)}_{\mathbb{R}^d}$ are
quasiinvariant under compactly-supported diffeomorphisms of
${\mathbb{R}^d}$.

\subsection{Configuration spaces of closed subsets}

A much larger configuration space, introduced in early work by
Ismagilov \cite{Ismag1971,Ismag1972,Ismag1975,Ismag1996}, is the
space $\Omega_M$ of all (non-empty) closed subsets of the manifold
$M$. For any closed set $C \in \Omega_M$, define the natural
action of a diffeomorphism $\phi \in$ {\it Diff\/}$^c(M)$ on
$\Omega_M$ by $\phi\, C = \{\,\phi(\mathbf{x})\,|\,\mathbf{x} \in
C\}$. Evidently $\phi\,C$ also belongs to $\Omega_M$, and we have
a (right) group action.

A $\sigma$-algebra for $\Omega_M$ is generated by the family of
sets in $\Omega_M$ consisting of all closed subsets of a given
closed set. Thus for $C \in \Omega_M$ (i.e., for $C \subseteq M$
closed), let $\Omega_{\,C} = \{C^{\,\prime} \in
\Omega_M\,|\,C^{\,\prime} \subseteq C\,\}$. Then let
$\mathcal{B}_{\,\Omega_M}$ be the smallest $\sigma$-algebra
containing the family of sets $\{\,\Omega_{\,C}\}_{\,C
\,\subseteq\, M\,\mathrm{closed}}\,$. This $\sigma$-algebra can
also be obtained as the algebra of Borel sets with respect to a
topology on $\Omega_M$, for which a subbase is the family of sets
$\{\,C\,|\,C \cap {\mathcal O} \neq \O\,\}_{\mathcal{O}
\,\subseteq\, M\,\mathrm{open}}\,$; i.e., the family of subsets of
$\Omega_M$ whose elements {\it meet\/} a given open set
$\mathcal{O} \subseteq M$.

Evidently any locally finite configuration $\gamma \in \Gamma_M$
is also a closed subset of $M$, so that in general we have
$\Gamma_M \subset \Omega_M$.

\subsection{Configuration spaces of generalized functions}

Another possibility is to work with the dual space
$\mathcal{D}^{\,\prime}(M)$, as suggested by the CURs of the
semidirect product group mentioned in Sec. I. That is, a
configuration $\gamma \in \mathcal{D}^{\,\prime}(M)$ is a
continuous, linear, real-valued functional on $\mathcal{D}(M)$---a
{\it distribution\/} or {\it generalized function\/} on M. This is
especially convenient for representing Eq. \!(\ref{repU}), as we
can immediately write $\langle \gamma,f \rangle$ for the value
taken by $\gamma$ on the function $f \in \mathcal{D}(M)$.

Diffeomorphisms act on $\mathcal{D}^{\,\prime}(M)$ by the dual to
their action on $\mathcal{D}(M)$; i.e., $\phi \gamma$ is defined
for $\gamma \in \mathcal{D}^{\,\prime}(M)$ by $\langle \phi
\gamma, f \rangle$ = $\langle \gamma, f \circ \phi \rangle$ for
all $f \in \mathcal{D}(M)$. [With this definition and our earlier
convention, we have $(\phi_1 \phi_2) \gamma = \phi_2 (\phi_1
\gamma)$, so that the group action is a right action as desired.]
A $\sigma$-algebra in $\mathcal{D}^{\,\prime}(M)$ may be built up
directly from cylinder sets with Borel base \cite{GV1964}, or
$\mathcal{D}^{\,\prime}(M)$ can be endowed with the weak dual
topology and measures constructed on the corresponding Borel
$\sigma$-algebra.

Evidently $\Gamma_M$, or more specifically
$\,\Gamma_{\mathbb{R}^d}\,$, may be identified naturally with a
subset of $\mathcal{D}^{\,\prime}(M)$, or
$\mathcal{D}^{\,\prime}({\mathbb R}^d)$, by the correspondence
\begin{equation}
\gamma \,\to\, \sum_{{\bf x} \in\, \gamma} \,\delta_{\,\bf x}\,,
\label{distribution}
\end{equation}
where $\delta_{\,\bf x} \in \mathcal{D}^{\,\prime}(M)$ is the
evaluation functional (i.e., the Dirac $\delta$-function) defined
by $\langle \delta_{\,\bf x}, f\rangle = f({\bf x}),\,{\bf x} \in
M$.

The so-called ``vague topology'' in $\Gamma_M$ is in fact the
topology that $\Gamma_M$ inherits from the weak dual topology in
$\mathcal{D}^{\,\prime}(M)$. While $\Gamma_M$ is not a linear
space, the larger space $\mathcal{D}^{\,\prime}(M)$ is. In
addition to linear combinations of evaluation functionals (with
possibly distinct real coefficients), $\mathcal{D}^{\,\prime}(M)$
contain other kinds of configurations of physical importance, that
do not belong to $\Gamma_M$ and in some cases are not easily
identified with elements of $\Omega_M$. For example,
configurations may include terms that are {\it derivatives\/} of
$\delta$-functions, as well as generalized functions with support
on embedded submanifolds of $M$.

\subsection{Configuration spaces of embeddings and immersions}

Still another characterization of a ``large'' space of
configurations in $M$ begins with some other manifold (or manifold
with boundary) $L$, together with a set of maps $\alpha: L \to M$
that obey some specified regularity and continuity properties (for
which there are numerous possible choices). Then we call $L$ the
{\it parameter space\/} for the corresponding class of
configurations, and $M$ the {\it target space.\/} When $\alpha$ is
{\it injective\/} (so that self-intersection of the image of $L$
in the target space is not permitted), we have a configuration
space of {\it embeddings,\/} while without any such restriction we
have a larger space of {\it immersions.\/}

We have at the outset the choice of considering {\it
parameterized\/} or {\it unparameterized\/} maps. A space of
parameterized $C^k$ immersions consists of mappings
$\,\alpha(\theta),\, \theta \in L$, that are $C^k$ for some fixed
integer $k \geq 0$. For $\phi \in$ {\it Diff\/}$^c(M)$, the
formula $[\phi \alpha] (\theta) = \phi(\alpha(\theta))$ (i.e.,
$\phi \alpha = \phi \circ \alpha$) defines the desired (right)
group action on the space of parameterized immersions. In
addition, the group {\it Diff\/}$(L)$ acts on the space of
immersions (as a left action) by {\it reparamaterization,\/} so
that for $\psi \in$ {\it Diff\/}$(L)$, $\psi: \alpha \to \alpha
\circ \psi$.

Then an unparameterized immersion is just the {\it image set\/} $K
= \,\alpha (L) \subset M$, where the parameterization of $K$ has
been disregarded. Alternatively, we can think of the
unparameterized immersion as an equivalence class of parameterized
immersions {\it modulo\/} reparamaterization. Note that the action
of {\it Diff\/}$^c(M)$ on the space of (parameterized or
unparameterized) immersions leaves the corresponding space of
embeddings invariant as a subset, and preserves the continuity
properties of configurations in the space.

If $L$ is the circle $S^1$, for instance, configurations are $C^k$
{\it loops\/} in $M$. The embeddings are the non-self-intersecting
loops. The action of the diffeomorphism group also respects the
knot class of the loop. If $L$ is the closed interval
$[\,0,2\pi\,]$, configurations are finite arcs in $M$. Further
possibilities include ribbons, tubes, or higher-dimensional
submanifolds of $M$.

The configuration space of unparameterized immersions of $L$ in
$M$ is a subset of the configuration space $\Omega_M$, invariant
(as a set) under the action of {\it Diff\/}$^c(M)$. This
description thus allows us to {\it refine\/} $\Omega_M$ as
sensitively as desired, according to the topological and
continuity properties of extended configurations.

For example, quantized vortex configurations in ideal,
incompressible fluids are obtained from representations of groups
of (area- and volume-preserving) diffeomorphisms of $\mathbb{R}^2$
and $\mathbb{R}^3$. For planar fluids, pure point vortices are not
permitted quantum-mechanically, but one-dimensional {\it
filaments\/} of vorticity are allowed. Similarly, in
$\mathbb{R}^3$ pure filaments are kinematically forbidden, while
two-dimensional vortex {\it surfaces\/}, e.g. ribbons or tubes,
can occur \cite{RasReg1975, MarWein1983, GolMenSh1987,
GolMenSh1991}. But a major gap is the construction of measures,
quasiinvariant under diffeomorphisms, directly on spaces of
filaments or tubes. One approach to the filament case has been
suggested by Shavgulidze \cite{Shav1999}.

Naturally a nonrelativistic quantum theory of strings, with
$\mathbb{R}^d$ as the target space, also depends on quasiinvariant
measures on the space of loops.

In addition, we remark that diffeomorphism-invariant measures are
important to the long-standing problem of finding consistent
theories for quantized gravity; for instance, Ashtekar and
Lewandowski have constructed a faithful, diffeomorphism-invariant
measure on a compactification of the space of gauge-equivalent
connections \cite{AshLew1994, MarMou1995}.

Reparamaterization invariance has nice consequences for quantum
mechanics, when expressed in terms of diffeomorphism group
representations. Note in particular that we can consider the
$N$-particle configuration space $\Gamma^{(N)}_M$ as a special
case of embeddings {\it modulo\/} reparamaterization, with the
discrete manifold $L = \{1, \dots , N\}$. The group {\it
Diff\/}$(L)$ in this case is the symmetric group $S_N$. The
corresponding configuration space of parameterized embeddings is
the space of {\it ordered\/} $N$-tuples $({\bf x}_1, \dots, {\bf
x}_N)$ of distinct points, ${\bf x}_j \neq {\bf x}_k$ for $j \neq
k$. The space of parameterized immersions of $L$ in $M$ includes
the $N$-tuples with coincident points.

\subsection{The configuration space of countable \\ subsets of $\mathbb{R}^d$}

The idea pursued in the balance of this article is the
construction of measures, quasiinvariant under diffeomorphisms of
$\mathbb{R}^d$, on the space $\Sigma_{\mathbb{R}^d}^{(\infty)}$ of
countably infinite subsets of the physical space $\mathbb{R}^d$
that are {\it not\/} necessarily locally finite. Alternatively, we
may work on the space $\Sigma_{\mathbb{R}^d}$ whose elements are
subsets $\gamma \subset \mathbb{R}^d$ that are finite or countably
infinite, with
\begin{equation}
\Sigma_{\mathbb{R}^d}\,=\, \bigsqcup_{\,N=1}^{\,\infty}
\Gamma_{\mathbb{R}^d}^{(N)} \bigsqcup
\Sigma^{(\infty)}_{\mathbb{R}^d}\,. \label{defineSigma}
\end{equation}
We call this the space of {\it generalized configurations.\/}

Our main mathematical motivation for working with this space is
that measures on it can be constructed by means of random point
processes on spaces of {\it infinite sequences\/} of points in
$\mathbb{R}^d$. We shall project the measure $\mu$ on
$[\mathbb{R}^d]^\infty$ that results from such a point process to
define the corresponding measure $\hat{\mu}$ on
$\Sigma_{\mathbb{R}^d}$, thus obtaining a measure on the space
$\Sigma_{\mathbb{R}^d}^{(\infty)}$.

A physical motivation for this direction of work is the goal of
constructing quasiinvariant measures for spatially-extended
systems, which is in general an unsolved problem. Since
$\mathbb{R}^d$ is separable, any closed set in $\mathbb{R}^d$ can
be obtained as the closure of an element of
$\Sigma_{\mathbb{R}^d}$; so that the {\it closure map\/} $\,\gamma
\to \bar{\gamma}\,$ from $\Sigma_{\mathbb{R}^d}$ to
$\Omega_{\mathbb{R}^d}$ is surjective. Thus our present
approach---which puts us into a still larger configuration space
than that of Ismagilov---may permit point-like approximations to
embedded manifold configurations.

Apart from this general consideration, the specific measures we
can construct appear to have a direct interpretation as
descriptive of idealized quantum or statistical configurations
forming ``particle clouds'' about a locus of condensation. These
allow for a kind of ``phase transition'' from a rarefied to a
condensed phase, as the self-similarity parameter passes through a
critical value.

Let us write $\omega = ({\bf x}_j) \in [\mathbb{R}^d]^\infty$ to
denote an infinite sequence, with $j = 1,2,3,\dots$.

Now generalized configurations, like infinite sequences, can have
accumulation points. A point ${\bf x} \in {\mathbb{R}^d}$ is an
accumulation point of a set $\gamma \subset \mathbb{R}^d$---or,
respectively, of an infinite sequence $\omega = ({\bf x}_{j}) \in
[{\mathbb{R}^d}]^{(\infty)}$---if for any neighborhood ${\mathcal
U}$ of ${\bf x}$, the set ${\mathcal U} - \{{\bf x}\}$ contains
infinitely many points of $\gamma$ (respectively, $\omega$). An
accumulation point of $\gamma$ may or may not itself be an element
of $\gamma$. Evidently diffeomorphisms of $\mathbb{R}^d$ act
naturally on generalized configurations, respecting accumulation
points: if ${\bf x} \in \mathbb{R}^d$ is an accumulation point of
$\gamma \in \Sigma_{\mathbb{R}^d}^{(\infty)}$, then $\phi({\bf
x})$ is an accumulation point of $\phi \gamma$. The points
belonging to configurations in $\Sigma_{\mathbb{R}^d}^{(\infty)}$
can cluster in such a manner as to yield fractals or even more
complicated objects.

The set of sequences containing coincident points is called the
''diagonal'' $D$ in $[\mathbb{R}^d]^\infty$; that is, $D = \{({\bf
x_j}) \in [\mathbb{R}^d]^\infty \,|\, {\bf x_k} = {\bf
x_\ell}\,(\mathrm{for \,some}\,k \neq \ell)\}$. Typically $D$ is
of measure zero for the point processes of interest, and for
technical reasons it will often be convenient to exclude it. We
have the natural projection from the sequence space to the
configuration space, $p: [\mathbb{R}^d]^\infty \to
\Sigma_{\mathbb{R}^d}$, given by $p\,[({\bf x}_j)] = \{{\bf
x}_j\}$. The image of $[{\mathbb{R}^d}]^{\infty}$ under $p$ is
{\it all\/} of $\Sigma_{\mathbb{R}^d}$, since the possibility of
repeated entries in elements of $[{\mathbb{R}^d}]^{\infty}$
permits the corresponding configurations to be finite as well as
infinite. Then $[{\mathbb{R}^d}]^\infty$ can also be thought of as
a fiber space over $\Sigma_{\mathbb{R}^d}$. It is natural to
consider also the restriction of $p$ to sequences without repeated
entries, $p: [\mathbb{R}^d]^\infty - D \to
\Sigma_{\mathbb{R}^d}^{(\infty)}$ (which is surjective).

Note that the space $\Sigma^{(\infty)}_{\mathbb{R}^d}$ may also be
regarded as a special case of the space of unparameterized
embeddings discussed in the preceding subsection. The target space
$M$ is $\mathbb{R}^d$; the parameter space $L$ is $\mathbb{N}$
(the set of natural numbers); and {\it Diff\/}$(L)$ is the group
$S^{\infty}$ of bijections of $\mathbb{N}$. Of course,
$[{\mathbb{R}^d}]^\infty - D$ is then seen as the space of
parameterized embeddings of $L$ into $M$; while
$[{\mathbb{R}^d}]^\infty$ itself is the space of parameterized
immersions.

For any diffeomorphism $\phi$ of $\mathbb{R}^d$, we have $\phi
p\,[({\bf x}_j)] = \{\phi({\bf x}_j)\} = p\,[(\phi({\bf x}_j))]$.
Thus we can project a probability measure on the sequence-space
$[\mathbb{R}^d]^\infty$ or $[\mathbb{R}^d]^\infty - D$,
constructed as is usual from an infinite sequence of conditional
probability densities on $\mathbb{R}^d$, to a probability measure
on the configuration space $\Sigma_{\mathbb{R}^d}^{(\infty)}$,
consistent with the action of {\it Diff\/}$^c (\mathbb{R}^d)$.

In earlier work, it was shown how for the one-dimensional
manifolds $\mathbb{R}^1$ or $S^1$, {\it self-similar\/} point
processes in the manifold lead quite generally through such a
construction to quasiinvariant measures on the configuration space
of countably infinite subsets \cite{GolMos1995a, GolMos1995b,
GolMos1995c, Gol1996, GolMos2001}. The quasiinvariance is
intimately related to the self-similarity. In Sec. IV we shall
discuss further the relevant $\sigma$-algebra on this
configuration space, which lays the foundation for completing the
rigorous proofs of earlier conjectures. Then we shall indicate how
the generalization to $d > 1$ is carried out \cite{Sak2002}.

\section{Topology and Measurable Structure on
$\Sigma_{\mathbb{R}^d}$}

There are at least two possible approaches to defining a
$\sigma$-algebra on the generalized configuration space
$\Sigma_{\mathbb{R}^d}^{(\infty)}$.

\subsection{Indirect approach through $[{\mathbb{R}^d}]^\infty$}

The indirect approach makes use of the sequence space
$[{\mathbb{R}^d}]^\infty$, which is endowed with the well-known
{\it weak product topology\/} $\tau_w$. Let us write ${\bf
x}_j(\omega)$ for the $jth$ entry of $\omega \in
[{\mathbb{R}^d}]^\infty$. The weak topology is then the coarsest
topology for which all the natural projections $\pi_j:
[\mathbb{R}^d]^{\infty} \to {\mathbb{R}^d}$ given by $\omega \to
{\bf x}_j(\omega)$ are continuous. This topology is inherited by
$[{\mathbb{R}^d}]^\infty - D$.

Let $\mathcal{B}(\,[{\mathbb{R}^d}]^\infty )$ denote the
$\sigma$-algebra of Borel sets in $[{\mathbb{R}^d}]^\infty$ with
respect to $\tau_w$. This naturally induces a $\sigma$-algebra in
$\Sigma_{\mathbb{R}^d}$---namely, the largest $\sigma$-algebra
with the property that the projection $p: [\mathbb{R}^d]^\infty
\to \Sigma_{\mathbb{R}^d}$ is measurable \cite{GolMos1995b,
GolMos1995c}. More precisely, we introduce in
${\Sigma_{\mathbb{R}^d}}$ the $\sigma$-algebra
\begin{equation}\label{siggm}
{\mathcal P}_w({\Sigma_{\mathbb{R}^d}}) := \{A \subseteq
{\Sigma_{\mathbb{R}^d}}\ | \ p^{-1}(A) \in
\mathcal{B}(\,[{\mathbb{R}^d}]^\infty)\}\,,
\end{equation}
to which each of the subsets $\Gamma_{\mathbb{R}^d}^{(N)}$, $N =
1,2,3,\dots$, as well as $\Sigma^{(\infty)}_{\mathbb{R}^d}$,
belongs.

Evidently the set of accumulation points of an infinite sequence
in ${\mathbb{R}^d}$ or ${\mathbb{R}^d} - D$ may be empty, finite
and non-empty, countably infinite, or uncountably infinite. Since
accumulation points in ${\mathbb{R}^d}$ depend only on the {\it
set\/} $\gamma = \{{\bf x}_j\}$, and not specifically on the {\it
sequence\/} $({\bf x}_j)$, all the distinct elements of
$p^{-1}(\gamma)$ have precisely the same accumulation points.

Now it is straightforward to demonstrate that various sets of
interest in ${\Sigma^{(\infty)}_{\mathbb{R}^d}}$ belong to
${\mathcal P}_w$, by showing that the corresponding sets of
sequences belong to ${\mathcal B}([{\mathbb{R}^d}]^\infty - D)$. A
series of lemmas in earlier work \cite{Gol2000,GolMos2000} shows
that the set $[{\mathbb{R}^d}]^\infty -D$ itself belongs to
${\mathcal B}([{\mathbb{R}^d}]^\infty)$, and that the following
subsets of $[{\mathbb{R}^d}]^\infty -D$ are likewise Borel: the
set of all nonrepeating sequences having precisely $n$ elements in
a given compact set $K \subset {\mathbb{R}^d}$; the set of all
locally finite nonrepeating sequences; and the set of all
nonrepeating sequences having precisely $N$ accumulation points in
$K$. Each of these sets is the inverse image in
$[{\mathbb{R}^d}]^\infty - D$ (under the projection $p$) of a set
in $\Sigma_{\mathbb{R}^d}^{(\infty)}$; hence the corresponding
sets in $\Sigma_{\mathbb{R}^d}^{(\infty)}$ are measurable.

In fact, ${\mathcal P}_w({\Sigma^{(\infty)}_{\mathbb{R}^d}})$ is
sufficiently rich to permit us to count the numbers of
accumulation points of configurations that are located in
arbitrary Borel sets of ${\mathbb{R}^d}$ (not just compact sets).
In particular, the subsets $\Sigma^{\,(\infty)}_{{\mathbb{R}^d},N}
\subset \Sigma^{\,(\infty)}_{\mathbb{R}^d}$ consisting of
generalized configurations having exactly $N$ accumulation points
in ${\mathbb{R}^d}$ are measurable. The inverse image
$p^{-1}(\Sigma^{\,(\infty)}_{{\mathbb{R}^d},N})$ is the set of
infinite sequences having precisely $N$ accumulation points, which
we denote by $[{\mathbb{R}^d}]^\infty_N \subset
[{\mathbb{R}^d}]^\infty$ (for $N = 0,1,2,\dots$).

Suppose that we have a probability measure $\,\mu\,$ on
$[{\mathbb{R}^d}]^{\infty}$ or $[{\mathbb{R}^d}]^{\infty}-D$. Then
we obtain a probability measure $\,\hat{\mu}\,$ on
$\Sigma_{\mathbb{R}^d}$ by defining, for all $A \in {\mathcal
P}_w(\,\Sigma_{\mathbb{R}^d})$, $\,\hat{\mu}\,(A)$ =
$\,\mu\,(p^{\,-1}\,(A))$. The most straightforward way to
construct a countably additive measure $\mu$ on
$[{\mathbb{R}^d}]^{\infty}$ [with the $\sigma$-algebra ${\mathcal
B}([{\mathbb{R}^d}]^{\infty})$] is to specify a compatible family
of measures on the finite-dimensional spaces from which
$[\mathbb{R}^d]^\infty$ is constructed as the projective limit.
The existence of the corresponding measure $\mu$ is then assured
by Kolmogorov's theorem. If $\mu$ is quasiinvariant under
diffeomorphisms of ${\mathbb{R}^d}$, then our construction ensures
that $\,\hat{\mu}$ is also quasiinvariant as desired.

\subsection{Direct approach}

The more direct approach to constructing a $\sigma$-algebra on
$\Sigma _{\mathbb{R}^d}^{(\infty)}$ is simply to specify a
generating set of subsets of $\Sigma_{\mathbb{R}^d}$ or
$\Sigma_{\mathbb{R}^d}^{(\infty)}$ for the $\sigma$-algebra, or
else to introduce a topology in $\Sigma_{\mathbb{R}^d}$ or
$\Sigma_{\mathbb{R}^d}^{(\infty)}$ and to take as our
$\sigma$-algebra the Borel sets with respect to that topology.

For instance, we may begin with Ismagilov's $\sigma$-algebra on
$\Omega_{\mathbb{R}^d}$ described above, and lift it to a
$\sigma$-algebra ${\mathcal I}\,(\Sigma_{\mathbb{R}^d})$ using the
closure map. The generating family for ${\mathcal
I}\,(\Sigma_{\mathbb{R}^d})$ becomes all sets of the form $\{
\gamma \in \Sigma_{\mathbb{R}^d} \,|\, \gamma \subseteq F \}$,
where $F \in \Omega_{\mathbb{R}^d}$ is closed. Because $F$ is
closed, $\gamma \subseteq F$ if and only if $\bar{\gamma}
\subseteq F$. The complement of a set in this generating family is
the set ${\mathcal O}_U = \{\gamma \in \Sigma_{\mathbb{R}^d} \, |
\ \gamma \cap U \neq \emptyset\}$, the set of all configurations
that {\it meet\/} the open set $U \subseteq \mathbb{R}^d$; where
$U$ is $\mathbb{R}^d - F$. The collection of sets $\{ {\mathcal
O}_U \,|\,U \subseteq \mathbb{R}^d \,\mathrm{open}\}$ likewise
serves as generating family for ${\mathcal
I}\,(\Sigma_{\mathbb{R}^d})$ \cite{Sak2002}. The subsets
$\Gamma_{\mathbb{R}^d}^{(N)}$ and
$\Sigma^{(\infty)}_{\mathbb{R}^d}\,$ of $\Sigma_{\mathbb{R}^d}$
belong to ${\mathcal I}\,(\Sigma_{\mathbb{R}^d})$.

We can make use of these families of sets to introduce a natural
topology on $\Sigma_{\mathbb{R}^d}$. Define a subbase of open sets
for a topology $\tau_o$ in $\Sigma_{\mathbb{R}^d}$ to be
$\{{\mathcal O}_U \, | \, U\subseteq {\mathbb{R}^d} \,\mathrm{
open}\}$. Note that for any index set $I$, $\cup_{\alpha \in I}
\,{\mathcal O}_{U_{\alpha}} \,=\, {\mathcal O}_{\,[\,\cup_{\alpha
\in I}\,U_\alpha\,]}$, while $\cap_{\,j \,=\, 1, \dots ,n}
\,{\mathcal O}_{U_j} \supset {\mathcal O}_{\,[\,\cap_{j \,=\, 1,
\dots ,n}\, U_j\,]}$. The finite intersections of sets in the
subbase form a base for $\tau_o$.

In the topology $\tau_o$, the subsets $\Gamma_{\mathbb{R}^d}^{(n)}
\subset \Sigma_{\mathbb{R}^d}$ (for $n > 1$) and
$\Sigma_{\mathbb{R}^d}^{(\infty)} \subset \Sigma_{\mathbb{R}^d}$
are neither open nor closed. However, for each $N \geq 0$,
$\{\gamma\,| \ |\gamma| \leq N \} = \bigsqcup_{\,n=1}^{\,N}
\Gamma_{\mathbb{R}^d}^{(n)}$ is closed. Of course, we may also
consider separately the topology induced in
$\Sigma_{\mathbb{R}^d}^{(\infty)}$ by $\tau_o$.

Now the $\sigma$-algebra ${\mathcal I}\,(\Sigma_{\mathbb{R}^d})$,
that we obtained by lifting Ismagilov's $\sigma$-algebra to
$\Sigma_{\mathbb{R}^d}$ by the inverse image of the closure map,
is precisely the Borel $\sigma$-algebra ${\mathcal
B}_o(\Sigma_{\mathbb{R}^d})$ with respect to the topology
$\tau_o$. Indeed, we noted already that the complement of
${\mathcal O}_U$ in $\Sigma_{\mathbb{R}^d}$ is just $\{\gamma \in
\Sigma_{\mathbb{R}^d} \, | \, \gamma \, \subseteq \,
{\mathbb{R}^d}-U\}$. Thus we have immediately that ${\mathcal B}_o
(\Sigma_{\mathbb{R}^d})$ contains ${\mathcal
I}\,(\Sigma_{\mathbb{R}^d})$, and the closure map is
$\tau_o$-Borel measurable with respect to Ismagilov's
$\sigma$-algebra on $\Omega_{\mathbb{R}^d}$. Conversely, let
$\{U_j\,|\,j = 1,2,3, \dots\}$ be a countable base for the
topology in ${\mathbb{R}^d}$. Then $\{{\mathcal O}_{U_j}\}$ is a
countable subbase for $\tau_o$, and the finite intersections of
such sets form a countable base for $\tau_o$ whose elements are
obtained directly from the generating family for ${\mathcal
I}\,(\Sigma_{\mathbb{R}^d})$. Hence ${\mathcal B}_o
(\Sigma_{\mathbb{R}^d}) = {\mathcal I}\,(\Sigma_{\mathbb{R}^d})$.

Sakuraba constructs and discusses a related topology $\tau_s$ on
$\Sigma_M$ (here $M = \mathbb{R}^d$), obtained as a quotient of
the product topology on the disjoint union of $M^n, n = 1,2,3,
\dots \,,$ and $M^\infty$ with respect to the symmetric groups
$S_n$ and the infinite symmetric group \cite{Sak2002}. In this
construction, the topology on $\Sigma_M$ is the sum of topologies
on the components $\Gamma_M^{(n)}$ and $\Sigma_M^{(\infty)}$; and
each of the subsets $\Gamma_M^{(n)}$ is {\it both\/} closed and
open. Restricted to each component, $\tau_s$ coincides with the
topology induced by $\tau_o$. Thus the family of Borel sets of
$\tau_s$ coincides with the family of Borel sets of $\tau_o$.

The fact that ${\mathcal I}\,(\Sigma_{\mathbb{R}^d})\,
\subset\,{\mathcal P}_w(\Sigma_{\mathbb{R}^d})$ is
straightforward: since
\begin{equation}
p^{-1}\left({\mathcal O}_U \right) = \bigcup_{j=1}^{\infty}\{ \,
\omega \in [{\mathbb{R}^d}]^\infty \ | \ {\bf x}_j(\omega) \in U\,
\}\,,
\end{equation}
the inverse image of ${\mathcal O}_U$ is open in the weak topology
of $[{\mathbb{R}^d}]^\infty$, and therefore ${\mathcal O}_U$
belongs to ${\mathcal P}_w(\Sigma_{\mathbb{R}^d})$. But ${\mathcal
I}\,(\Sigma_{\mathbb{R}^d})$ is in fact smaller than $\,{\mathcal
P}_w(\Sigma_{\mathbb{R}^d})$, and too small for certain purposes.
Indeed, by our previous result any $\tau_o$-Borel set $B$ is the
inverse image under the closure map of a set in the
$\sigma$-algebra on $\Omega_{\mathbb{R}^d}$ generated by the sets
$\Omega_F$; thus it has the property that if $\gamma \in B$,
$\bar{\gamma} \in B$.

But it is easy to construct sets in ${\mathcal
P}_w(\Sigma_{\mathbb{R}^d})$ that do not have this property. For
example, define the set ${\mathcal O}^{\,V}$ of all configurations
$\gamma \in \Sigma_{\mathbb{R}^d}$ that are subsets of a given
open set $V$. Evidently, there exist countably infinite subsets of
$V$ whose closures are no longer subsets of $V$, so ${\mathcal
O}^{\,V}$ does not belong to ${\mathcal
I}\,(\Sigma_{\mathbb{R}^d})$. However ${\mathcal O}^{\,V}$ does
belong to ${\mathcal P}_w(\Sigma_{\mathbb{R}^d})$, which follows
from the fact that
\[
p^{-1}({\mathcal O}^{\,V}) \,=\,p^{-1}(\{\gamma\in
\Sigma_{\mathbb{R}^d} \, | \, \gamma \subset V\}) \]
\begin{equation}
= \bigcap_{j=1}^{\infty} \{\omega \mid {\bf x}_j(\omega) \in
V\}\,. \label{counterexample}
\end{equation}
Thus ${\mathcal I}\,(\Sigma_{\mathbb{R}^d}^{(\infty)})\, \neq
\,{\mathcal P}_w(\Sigma_{\mathbb{R}^d}^{(\infty)})$. The
$\sigma$-algebra ${\mathcal I}\,(\Sigma_{\mathbb{R}^d})$ is just
not large enough for us to be able to count the number of points
in a configuration that belong to a given open set in
$\mathbb{R}^d$.

This example suggests consideration of the {\it Vietoris
topology\/} on subsets of $\mathbb{R}^d$, restricted to
${\Sigma_{\mathbb{R}^d}}$ or to
$\Sigma_{\mathbb{R}^d}^{(\infty)}$. Let us call this topology
$\tau_v$. A subbase for $\tau_v$ is given by sets of the form
${\mathcal O}^{\,V} \cap {\mathcal O}_U$ where $U$ and $V$ are
open; so that ${\mathcal O}^{\,V}$ is itself open in $\tau_v$. The
Vietoris topology has many nice properties \cite{Viet1922,
Kur1966, AlipBor1999}. Considering then the $\sigma$-algebra
${\mathcal B}_v (\Sigma_{\mathbb{R}^d})$ of Borel sets with
respect to $\tau_v$, we have ${\mathcal
I}\,(\Sigma_{\mathbb{R}^d}) \subset {\mathcal B}_v
({\Sigma_{\mathbb{R}^d}})$, but ${\mathcal
I}\,(\Sigma_{\mathbb{R}^d}) \neq {\mathcal B}_v
({\Sigma_{\mathbb{R}^d}})$. Furthermore, ${\mathcal B}_v
({\Sigma_{\mathbb{R}^d}}) \subset {\mathcal
P}_w(\Sigma_{\mathbb{R}^d})$. To show
 this, consider again a countable base $\{U_j, \,\,j =
1,2,3,\dots\}$ for the topology in $\mathbb{R}^d$. A countable
subbase for $\tau_v$ is then $\{\,{\mathcal O}^{\,U_j} \cap
{\mathcal O}_{U_k}\,,\, j,k = 1,2,3, \dots \,\}$; and a countable
base for $\tau_v$ consists of finite intersections of such sets.
Since $p^{-1}({\mathcal O}^{\,U_j})$ and $p^{-1}({\mathcal
O}_{U_k})$ are both Borel in $[{\mathbb{R}^d}]^\infty$, the
inverse image of any open set in $\tau_v$ is Borel in
$[{\mathbb{R}^d}]^\infty$, which suffices for the result.

We have not, however, determined whether ${\mathcal
B}_v(\Sigma_{\mathbb{R}^d})$ is or is not strictly smaller than
$\,{\mathcal P}_w(\Sigma_{\mathbb{R}^d})$.

\section{Self-Similar Random Point Processes in $\mathbb{R}^d$
and Quasiinvariant Measures}

Now we are prepared to construct measures on the $\sigma$-algebra
$\mathcal{B}(\,[{\mathbb{R}^d}]^\infty)$ by means of random point
processes, using sequences of conditional probability densities.
When we do so, it turns out that the RN derivatives under
transformations by diffeomorphisms take the form of an infinite
product,
\begin{equation}
{{{d\mu_{\phi}} \over {d\mu}}(\omega)}=\prod_{j=1}^{\infty}u_{j,
\phi}(\omega)\,. \label{rn}
\end{equation}
Here $\omega \in [{\mathbb{R}^d}]^\infty$, and the $u_{j,
\phi}(\omega)$ are measurable functions that depend only on the
first $j$ entries of $\omega$.

Quasiinvariance of $\mu$ then requires that (\ref{rn}) converge to
a non-zero, non-infinite limit almost everywhere in $\mu$, for
each $\phi$. This means that the individual terms $u_{j,
\phi}(\omega)$ must approach $1$ sufficiently rapidly, as $j \to
\infty$. Under conditions that in fact hold for the measures
discussed here, these convergence properties have also been proven
sufficient to ensure the quasiinvariance of $\mu$ \cite{Sak2002},
and as a direct consequence, the quasiinvariance of the projected
measure $\hat{\mu}$ on $\Sigma _{\mathbb{R}^d}^{(\infty)}$.

Let $f({\bf x}_{j}|{\bf x}_{1},\dots,{\bf x}_{j-1})$ be a
non-singular probability density on $\mathbb{R}^d$ for selection
of the point ${\bf x}_j$, conditioned on the previously-selected
points ${\bf x}_{1},\dots,{\bf x}_{j-1}$ in some random sequence.
Then $d\mu_{j}({\bf x}_{j})=f({\bf x}_{j}|{\bf x}_{1},\dots,{\bf
x}_{j-1})d{\bf x}_{j}$ defines a conditional (Borel) probability
measure $\mu_{j}$ on ${\mathbb R}^d$ that depends measurably on
the $j-1$ real parameters ${\bf x}_{1},\dots,{\bf x}_{j-1}$ (the
positions of the first $j-1$ particle coordinates), and is
absolutely continuous with respect to the Lebesgue measure $d{\bf
x}_j\,$. We can interpret the joint probability measure for the
first $k$ points, specified by
$d\mu^{(k)}=\prod_{\,j=1}^{\,k}d\mu_{j}$, as a measure on
$[\mathbb{R}^d]^\infty$; and the sequence $(\mu^{(k)}), \, k =
1,2,3, \dots$, is then a compatible family of probability
measures.

By Kolmogorov's theorem, there is a unique measure $\mu$ on
$[\mathbb{R}^d]^\infty$ determined by the sequence $(\mu^{(k)})$.
Under transformation by $\phi \in$ {\it
Diff\/}$^c(\mathbb{R}^{d})$, the RN derivative for $\mu^{(k)}$
(when it exists) is given by the finite product
\begin{equation}
\frac{d\mu^{(k)}_{\phi}}{d\mu^{(k)}}(\omega) = \prod_{j=1}^{k}
\frac{d\mu_{j,\phi}}{d\mu_{j}}(\omega)\,, \label{rnf}
\end{equation}
where
\begin{equation}
\frac{d\mu_{j,\phi}}{d\mu_{j}}(\omega) = \frac{f(\phi({\bf
x}_{j})|\phi({\bf x}_{1}),\ldots,\phi({\bf x}_{j-1}))} {f({\bf
x}_{j}|{\bf x}_{1},\ldots,{\bf x}_{j-1})}{\mathcal J}_{\phi}({\bf
x}_{j}). \label{RNformula}
\end{equation}
The quasiinvariance of $\mu^{(k)}$ is assured as long as the RN
derivative in Eq. (\ref{rnf}) is almost everywhere positive and
finite. Now, as anticipated, in the infinite-dimensional case
quasiinvariance of the measure $\mu$ under diffeomorphisms turns
out to depend on the behavior of the infinite product in Eq.
\!(\ref{rn}), with $u_{j,\phi}({\omega})=
[d\mu_{j,\phi}/d\mu_{j}](\omega)$.

Of course, not every measure so constructed will be
quasiinvariant. The idea that leads to an interesting class of
quasiinvariant measures is to {\it scale\/} the probability
distribution of the $j$th particle's position according to the
{\it outcomes\/} for the previously chosen particle positions.
This establishes a self-similar random process, where in the
vicinity of accumulation points the ratio of probability density
functions in Eq. \!(\ref{RNformula}) approaches the inverse of the
Jacobian as $j \to \infty$. The resulting physical systems behave
like an interacting gas of particles with one or more loci of
condensation. However, our approach differs from the usual one in
that our probability measures are constructed directly, rather
than by means of an interaction Hamiltonian.

In general, if the positions of the particle coordinates ${\bf
x}_j(\omega)$, or the successive difference coordinates ${\bf
y}_{j+1}(\omega) = {\bf x}_{j+1}(\omega) - {\bf x}_j (\omega)$,
distribute independently but non-identically---so that points can
accumulate with non-zero probability---the resulting measure will
not be quasiinvariant. However, Ismagilov did demonstrate
quasiinvariance under diffeomorphisms of the measures resulting
from a particular class of processes of this type, in one space
dimension \cite{Ismag1971}.

Sakuraba \cite{Sak2002} showed that the quasiinvariant measures
constructed by Goldin and Moschella from self-similar random
processes, and the quasiinvariant measures of Ismagilov, are
mutually singular.

\subsection{Example for $d = 1$}

Let us illustrate with the examples based on Gaussian probability
densities. Working first with $d = 1$, choose an initial point
$x_0$ from a nowhere vanishing probability density $f_0$ on
$\mathbb{R}$. For $j = 1,2,3,\dots$, let $x_j = x_{j-1} + y_j$,
where the $y_j$ are a sequence of deviation values. Choose the
value $y_1$ from a unit normal distribution $g_1$, with mean $0$.
Given the values $y_1,\dots,y_j$, choose $y_{j+1}$ from a normal
distribution with mean $0$, and standard deviation $\sigma_j =
\kappa \, |y_j|$, where $\kappa > 0$ is a fixed {\it correlation
parameter\/} independent of $j$. Small values of $\kappa$
correspond to more tightly bound systems. Thus we have the
conditional probability densities for the $y_j$,
\begin{equation}
g_{j+1}^{\kappa}(y_{j+1}\,|\,y_{j})
=\frac{(2\pi)^{-\frac{1}{2}}}{\kappa|y_{j}|}
\;\,\exp\left[-\frac{1}{2\kappa^{2}}
\left(\frac{y_{j+1}}{y_{j}}\right)^{2}\right]\,. \label{gaussian}
\end{equation}
For sufficiently small values of $\kappa$, $(y_j)$ converges to
$0$ (with probability one), while $\sum_{\,j=1}^{\,\infty} |y_j| <
\infty$.

Let {\it Diff\/}$^c_0 (\mathbb{R})$ denote the stability subgroup
of {\it Diff\/}$^c (\mathbb{R})$, consisting of the compactly
supported diffeomorphisms of $\mathbb{R}$ that leave the origin
fixed. The measure on the space of sequences $(y_j)$ resulting
from the densities in Eq. \!(\ref{gaussian}) is then
quasiinvariant under the action of elements of {\it Diff\/}$^c_0
(\mathbb{R})$. We thus obtain the random sequence $\omega =
(x_k)$, with $x_k = x_0 + \sum_{\,j=1}^{\,k} y_j$, and the
corresponding random configuration $\gamma = \{x_k\}$.

Defining the terms $u_{j,\phi}$ in Eq. \!(\ref{rn}) accordingly,
we obtain $u_j \to 1$ sufficiently rapidly to ensure convergence
of the infinite product. More precisely, there exists a critical
value $\kappa_{0}$ such that if $0 < \kappa < \kappa_{0}$,
sequences $(x_j)$ converge to an accumulation point with
probability one, while if $\kappa_0 < \kappa$, sequences diverge
geometrically with probability one. In both cases, the associated
measures on $\Sigma_{\mathbb{R}}^\infty$ are quasiinvariant under
compactly supported diffeomorphism of $\mathbb{R}$
\cite{GolMos1995a, GolMos1995b, GolMos1995c, Sak2002}. The proofs
make use of the strong law of large numbers.

The above is not tied essentially to the use of normal
distributions; all that is really necessary for is the scaling
property. Thus, for a whole class of models, there exists a
critical value $\kappa_0$ of the scaling parameter $\kappa$. For
$0 < \kappa < \kappa_0$, the generalized configuration $\{x_j\}$
has an accumulation point with probability one; we call this the
{\it condensed phase.\/} For $\kappa_0 < \kappa$, $\{x_j\}$ has
zero average density; we call this the {\it rarefied phase.\/} For
each value of $\kappa$ (except for the critical value itself), we
have a bona fide unitary representation of {\it
Diff\/}$^c({\mathbb R})$, describing the associated quantum
system.

\subsection{Generalization to $d > 1$}

It was suggested earlier that a procedure similar to that
suggested by Eq. \!(\ref{gaussian}) would work in $d$ space
dimensions, $d > 1$, to yield measures on the space of generalized
configurations quasiinvariant under {\it Diff\/}$^c({\mathbb
R}^d)$; with the conditional probability density for ${\bf
y}_{j+1}$ dependent on the preceding $d$ outcomes $({\bf
y}_{j-d+1}, \dots , {\bf y}_{j})$ through the covariance matrix of
a multivariate normal distribution \cite{GolMos1995b,
GolMos1995c}. The generalization obtained by Sakuraba
\cite{Sak2002} achieves this, but also involves some new aspects.

Consider a random process where, at each stage, $d$ vectors in
$\mathbb{R}^d$ are to be selected. Thus at each stage we are
choosing a $d \times d$ random matrix $V$, and it is appropriate
to think of $\omega \in [\mathbb{R}^d]^\infty$ as the sequence of
square matrices $([{\bf x}_1, \dots {\bf x}_d], [{\bf x}_{d+1},
\dots {\bf x}_{2d}], \dots)$.

For the square matrix $Y = [y_{ij}]$, define the norm $||Y|| =
[\,{\sum_{\,i,j=1}^{\,d} y_{ij}^2}\,]^{1/2}$. Note that $||Y||$ is
a vector norm, not the operator norm of the matrix. For $Y \in
GL(d, \mathbb{R})$, define the {\it condition number\/} $k\,(Y) =
||Y||\cdot ||Y^{-1}||$. We may write $Y = P\,|Y|$, where $P$ is an
orthogonal matrix and where $\,|Y| = \sqrt{Y^t Y}$ is positive.
Let $\tau_1, \dots, \tau_d$ be eigenvalues of the matrix $\,|Y|$.
Then $||Y|| = ||\,|Y|\,||,$ and
\[
||Y|| = [\,\Sigma_{\,i=1}^{\,d}\,\tau_i^{\,2}\,]^{1/2}, \quad
||Y^{-1}|| = [\,\Sigma_{\,j=1}^{\,d}\,\tau_j^{-2}\,]^{1/2},
\]
\begin{equation}
k\,(Y)= [\,\Sigma_{\,i,j=1}^{\,d}\,(\tau_i/\tau_j)^{2}\,]^{1/2}.
\label{kformula} \end{equation}
Evidently $k\,(Y)$ characterizes the amount of deformation under
linear transformation by $Y$. If $Y$ is not invertible, then
$k\,(Y)$ is undefined (or infinite). Such matrices belong to
measure zero sets in the constructions that follow.

We next construct a measure on $[\,GL(d, \mathbb{R})]^\infty$ and
thus on $[\mathbb{R}^d]^\infty$, quasiinvariant under {\it
Diff\/}$^c_0 (\mathbb{R}^d)$. Define the probability density
function
\begin{equation}
f(Y) = C \exp
\,\{\,-\frac{1}{2\kappa^2}\,[\,||Y||\,k\,(Y)\,]^2\,\}\,
\label{probdens}
\end{equation}
on the set of $d \times d$ matrices, where $C$ is a normalization
constant chosen so that $\int f(Y)\,dY = 1$; here $dY = d{\bf y}_1
\dots d{\bf y}_d$. Let $\mu^{(k)}$ be the probability measure
defined by
\begin{equation}
d\mu^{(k)} = f(Y_1)\frac{f(Y_1^{-1}Y_2)}{|\det Y_1|^d}\,\cdots\,
\frac{f(Y_{k-1}^{-1}Y_k)}{|\det Y_{k-1}|^d}\,dY_1\cdots dY_k\,,
\end{equation}
where $d\mu^{(k)} = d\mu^{(k)}(Y_1, \dots, Y_k)$. Then $\mu^{(k)}$
is concentrated on $[\,GL(d, \mathbb{R})]^k$; i.e., the set of
sequences with one or more non-invertible matrices is of measure
zero.

Then we again have a critical value $\kappa_0$. For $\kappa <
\kappa_0$, the sequence $(Y_j)$ of matrices---and thus the
sequence of component vectors $({\bf y}_i)$---converges to $0$
with probability one; while for $\kappa_0 < \kappa$, it diverges
with probability one. Furthermore, the projective limit measure
$\mu$ on $[\,GL(d, \mathbb{R})]^\infty$ has the desired property
of quasiinvariance under {\it Diff\/}$^{\,c}_0 (\mathbb{R}^d)$.
The presence of the condition number $k\,(Y)$ in Eq.
\!(\ref{probdens}) is essential for the estimates required in
demonstrating convergence of the infinite product in the resulting
expression for the RN derivative. The proof here again uses the
strong law of large numbers.

Eq. \!(\ref{probdens}) can be generalized, replacing $k\,(Y)$ by
$k\,(Y)^\alpha$ ($\alpha \geq 1$), and replacing the Gaussian
density by a more general probability density function.

Finally, we may begin with a matrix of positions $X_0 = [\,{\bf
x}_1, \dots {\bf x}_d\,]$, chosen from a nowhere vanishing
probability density. Let $\bar{\bf x}_0$ be the center of position
of the $d$ vectors comprising $X_0$. Now we may treat each new
matrix $Y_j$ as a set of deviations from the center of position of
the preceding set of vectors $X_{j-1}$; so that with obvious
notation, $X_j = \bar{\bf x}_{j-1} + Y_j$. In this manner, we
obtain a measure on $[\mathbb{R}^d]^\infty$ quasiinvariant under
{\it Diff\/}$^c (\mathbb{R}^d)$, that projects to a quasiinvariant
measure on the space $\Sigma_{\mathbb{R}^d}^\infty$ of generalized
configurations.

More details about the preceding results may be found in the
thesis of Sakuraba \cite{Sak2002}, and in forthcoming
publications.

\section{Conclusion}

We believe the work summarized here strengthens the case for
basing a theory of statistical physics in the manifold $M$ on the
configuration space $\Sigma_M$ of countable subsets of $M$,
endowed with the Vietoris topology. Measures obtained from random
point processes in $M$ project to measures on $\Sigma_M$, and when
we consider self-similar random processes, we obtain measures
quasiinvariant under the group of compactly-supported
diffeomorphisms of $M$. The problem of relating these measures to
Hamiltonians on a classical phase space remains open.

\begin{acknowledgments}

G. Goldin wishes to thank Universit\`a dell'Insubria  and INFN,
sez. di Milano, for hospitality during his recent visits, and to
acknowledge the support of the Alexander von Humboldt Foundation
for related research.

\end{acknowledgments}

\end{document}